\documentclass[twocolumn,secnumarabic,amssymb,showpacs,superscriptaddress, aps, prl]{revtex4-1}

\usepackage{amsmath} 
\usepackage{amsthm} 
\usepackage{amssymb}	
\usepackage{graphicx} 
\usepackage[table]{xcolor}
\usepackage{comment}
\usepackage[export]{adjustbox}
\usepackage{hyperref}
\hypersetup{
	colorlinks  = true,
	citecolor    = blue
}
\usepackage{blindtext}
\usepackage{enumitem}
\usepackage{lipsum}

\def\beq{\begin{equation}}
\def\eeq{\end{equation}}
\def\bea{\begin{eqnarray}}
\def\eea{\end{eqnarray}}

\newcommand{\ket}[1]{\left| #1 \right>} 

\begin{document}

\title{Depolarization dynamics in a strongly interacting solid-state spin ensemble}
\author{J. Choi}
\thanks{These authors contributed equally to this work}
\affiliation{Department of Physics, Harvard University, Cambridge, Massachusetts 02138, USA}
\affiliation{Harvard John A. Paulson School of Engineering and Applied Sciences, Harvard University, Cambridge, MA 02138, USA}
\author{S. Choi}
\thanks{These authors contributed equally to this work}
\affiliation{Department of Physics, Harvard University, Cambridge, Massachusetts 02138, USA}  
\author{G. Kucsko}
\thanks{These authors contributed equally to this work}
\affiliation{Department of Physics, Harvard University, Cambridge, Massachusetts 02138, USA} 
\author{P. C. Maurer}
\affiliation{Department of Physics, Stanford University, Stanford, California 94305, USA}
\author{B. J. Shields}
\affiliation{Department of Physics, Harvard University, Cambridge, Massachusetts 02138, USA} 
\author{H. Sumiya}
\affiliation{Sumitomo Electric Industries Ltd., Itami, Hyougo, 664-0016, Japan}
\author{S. Onoda}
\affiliation{Takasaki Advanced Radiation Research Institute, National Institutes for Quantum and Radiological Science and Technology, 1233 Watanuki, Takasaki, Gunma 370-1292, Japan}
\author{J. Isoya}
\affiliation{Research Centre for Knowledge Communities, University of Tsukuba, Tsukuba, Ibaraki 305-8550, Japan}
\author{E. Demler}
\affiliation{Department of Physics, Harvard University, Cambridge, Massachusetts 02138, USA} 
\author{F. Jelezko}
\affiliation{Institut f{\"u}r Quantenoptik, Universit{\"a}t Ulm, 89081 Ulm, Germany}
\author{N. Y. Yao}
\affiliation{Department of Physics, University of California Berkeley, Berkeley, California 94720, USA.}
\author{M. D. Lukin}
\affiliation{Department of Physics, Harvard University, Cambridge, Massachusetts 02138, USA}

\begin{abstract}
We study the depolarization dynamics of a dense ensemble of dipolar interacting spins, associated with nitrogen-vacancy centers in diamond. We observe anomalously fast, density-dependent, and non-exponential spin relaxation.
To explain these observations, we propose a microscopic model where an interplay of long-range interactions, disorder, and dissipation leads to predictions that are in quantitative agreement with both current and prior experimental results. 
Our results pave the way for controlled many-body experiments with long-lived and strongly interacting ensembles of solid-state spins.
\end{abstract}

\maketitle

Electronic spins associated with solid-state point defects are promising candidates for the realization of quantum bits and their novel applications~\cite{childress2014atom,dolde2013room,michalet2005quantum,lovchinsky2016nuclear,amsuss2011cavity}.
In particular, the precise quantum control of {\it individual} nitrogen-vacancy (NV) centers in diamond has led to advances in both fundamental physics~\cite{waldherr2011violation,pfaff2013demonstration,hensen2015loophole} and the development of applications ranging from nanoscale sensing to quantum information science~\cite{maze2008nanoscale,kucsko2013nanometre,mamin2013nanoscale,lovchinsky2016nuclear,maurer2012room,yao2011robust,yao2012scalable}.
This high degree of control naturally suggests the use of strongly interacting, {\it dense} NV ensembles to explore quantum many-body dynamics~\cite{kucsko2016critical}. 

However, a key challenge in this context is the apparent reduction of the electronic spin lifetime at high defect densities~\cite{budker2015shortT1,jarmola2015longitudinal,kucsko2016critical}.
Such effects were first observed in  phosphorus doped silicon over five decades ago~\cite{feher1959electron}, where it was suggested that anomalously fast spin relaxation could arise from electronic hopping between nearby impurities. 
In addition to reduced spin lifetimes, recent experiments in dense NV ensembles have also  observed that this relaxation is relatively insensitive to temperature over a wide range, implying that the underlying mechanism is qualitatively different from the phonon-induced depolarization of single, isolated NV centers~\cite{budker2015shortT1,jarmola2015longitudinal,shrivastava1983theory}.

In this Letter, we characterize the depolarization dynamics of high-density NV ensembles at room temperature.
In particular, we perform spin lifetime measurements under different conditions, varying initial state populations, resonant spin densities, and microwave driving strength. 
To explain the observed features in the spin dynamics, we introduce a spin-fluctuator model, in which a network of short-lived spins (fluctuators) causes depolarization of nearby NV centers via dipolar interactions.
Moreover, additional measurements reveal the presence of charge dynamics, providing a potential microscopic origin for such fluctuators.

\begin{figure}[t]
\includegraphics[width=0.48\textwidth]{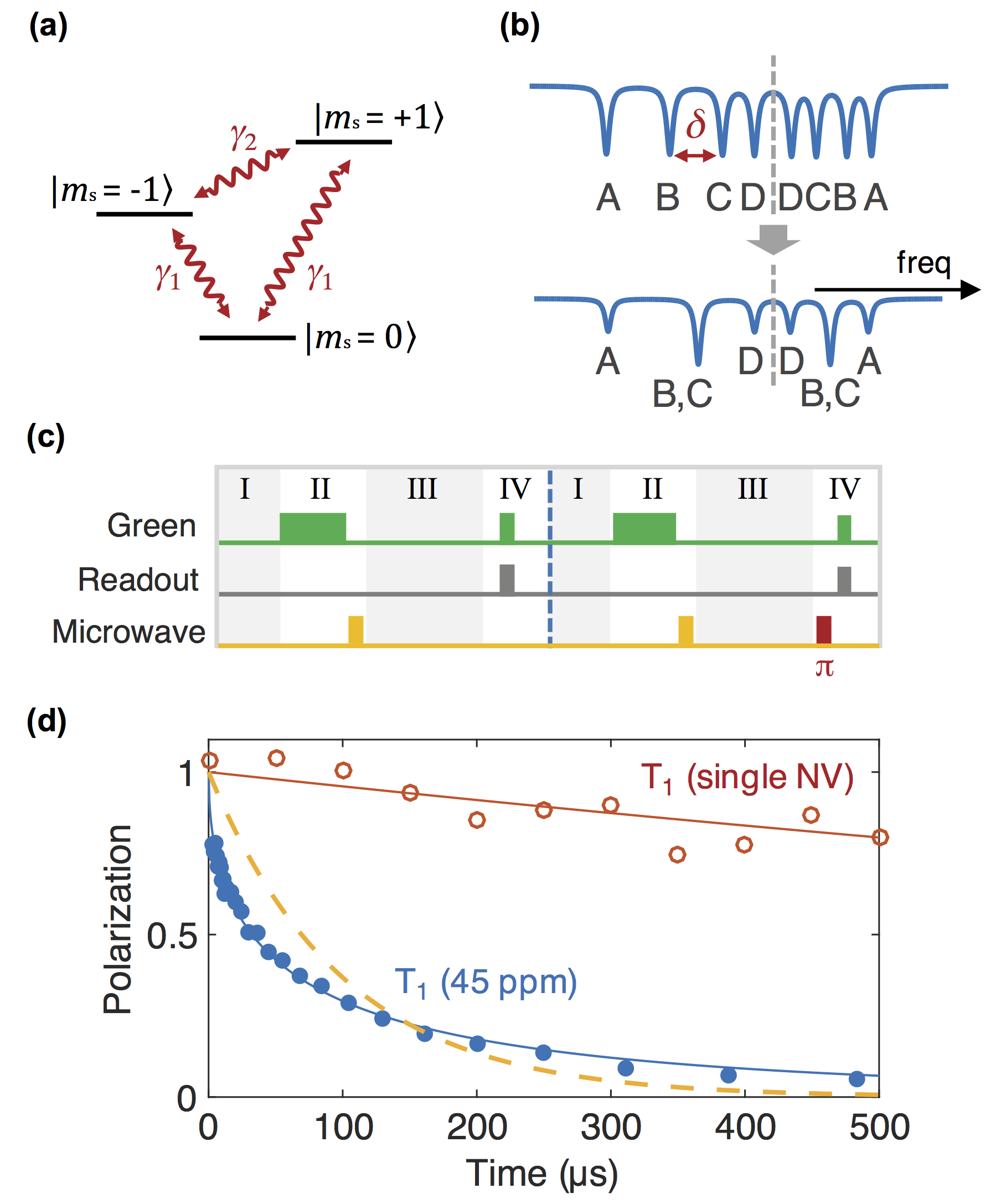} 
\caption{\textbf{Experimental System.} 
\textbf{(a)} Level diagram for NV centers. The degeneracy of the $\ket{m_s = \pm 1}$ spin states can be lifted by an external magnetic field. Red arrows indicate possible spin decay channels, $\gamma_1$ and $\gamma_2$.
\textbf{(b)} Schematic electron spin resonance (ESR) spectrum of four groups of NV centers {A,B,C,D} with spectral separation $\delta$ between B and C (upper curve). The effective density of resonant spins can be tuned by changing the orientation of the external magnetic field (lower curve).
\textbf{(c)}  Experimental sequence used to measure NV dynamics. I: charge equilibration ($\sim100$~$\mu$s duration); II: spin polarization (10~$\mu$W laser power, $\sim200$~$\mu$s duration) and subsequent microwave manipulation to modify the initial state; III: time evolution; IV: spin state readout. Red $\pi$-pulse is used to measure the population in an orthogonal state.
\textbf{(d)} Comparison of depolarization timescales between a single NV (red data, exponential fit) and a dense ensemble of NVs ($\sim45$~ppm, blue data, stretched-exponential fit with $T_1 \sim 67 ~\mu s$). The dashed line is a simple exponential curve with a time constant of 100~$\mu$s for comparison. 
}
\end{figure}

Our sample is fabricated from a type-Ib HPHT single crystal diamond, irradiated with an electron beam to create vacancies. A high concentration of NV centers was achieved via high fluence and \emph{in situ} annealing to avoid degradation of the crystal lattice. The resulting sample contains $\sim45$~ppm of NV centers, corresponding to a typical dipolar interaction strength, $J\sim (2\pi)~420$~kHz, significantly faster than extrinsic decoherence rates. To achieve a high degree of spatial control over the optical excitation region, a diamond nanobeam (300 nm x 300 nm x 20um) is created via angle-etching and used for all experiments in this work unless otherwise noted~\cite{kucsko2016critical}.

Each NV center constitutes an effective spin-1 system, which can be optically initialized, manipulated, and read out at ambient conditions~\cite{manson2006nitrogen}.
In the absence of an external magnetic field, the spin states $\ket{m_s=\pm1}$ are separated from the $\ket{m_s = 0}$ state by a crystal field splitting $\Delta_0 = (2\pi)~2.87$~GHz.
Applying a magnetic field further splits the $\ket{m_s=\pm1}$ states via a Zeeman shift,
which is proportional to the projection of the field onto the NV quantization axis (Fig.~1a).
Since NVs can be oriented along any of the four crystallographic axes of the diamond lattice, we can spectrally separate four groups of NV centers $\{A, B, C, D\}$ and independently control them via coherent microwave driving (Fig.~1b).
By tuning the direction of the magnetic field, we can additionally tune the number of spectrally overlapping groups and hence the effective density of spins.

{\it Experiments}---To probe the depolarization dynamics of strongly interacting NV ensembles,  we utilize the pulse sequence illustrated in Fig.~1c.
This sequence allows one to prepare and measure the population in an arbitrary spin state.
By repeating a specific sequence with an additional $\pi$-pulse (right panel, Fig.~1c), one measures the population of an orthogonal spin state and can use the difference between the two measurements, $P(t)$, to extract the depolarization dynamics~\cite{budker2015shortT1}.

To begin, we measure the depolarization time for a single group of NV centers (Fig.~1d).
The observed decay time $T_1 \lesssim 100 ~\mu s$ is significantly reduced when compared to isolated NVs, where typical lifetimes reach several milliseconds at room temperature~\cite{redman1991spin, walker19685,jarmola2012temperature}.
Moreover, the decay profile deviates from a simple exponential. Phenomenologically we find that it is characterized by a stretched exponential  with exponent 1/2
 \begin{align}
    P(t) = e^{-\sqrt{t/T_1}} \label{eq:stret},
\end{align}
%
consistent with several previous observations~\cite{jarmola2012temperature,jarmola2015longitudinal,budker2015shortT1}.
At differing spatial locations,  the extracted $T_1$ exhibits small variations possibly due to an inhomogeneous NV concentration.
\begin{figure}[t]
\includegraphics[width=.48\textwidth,left]{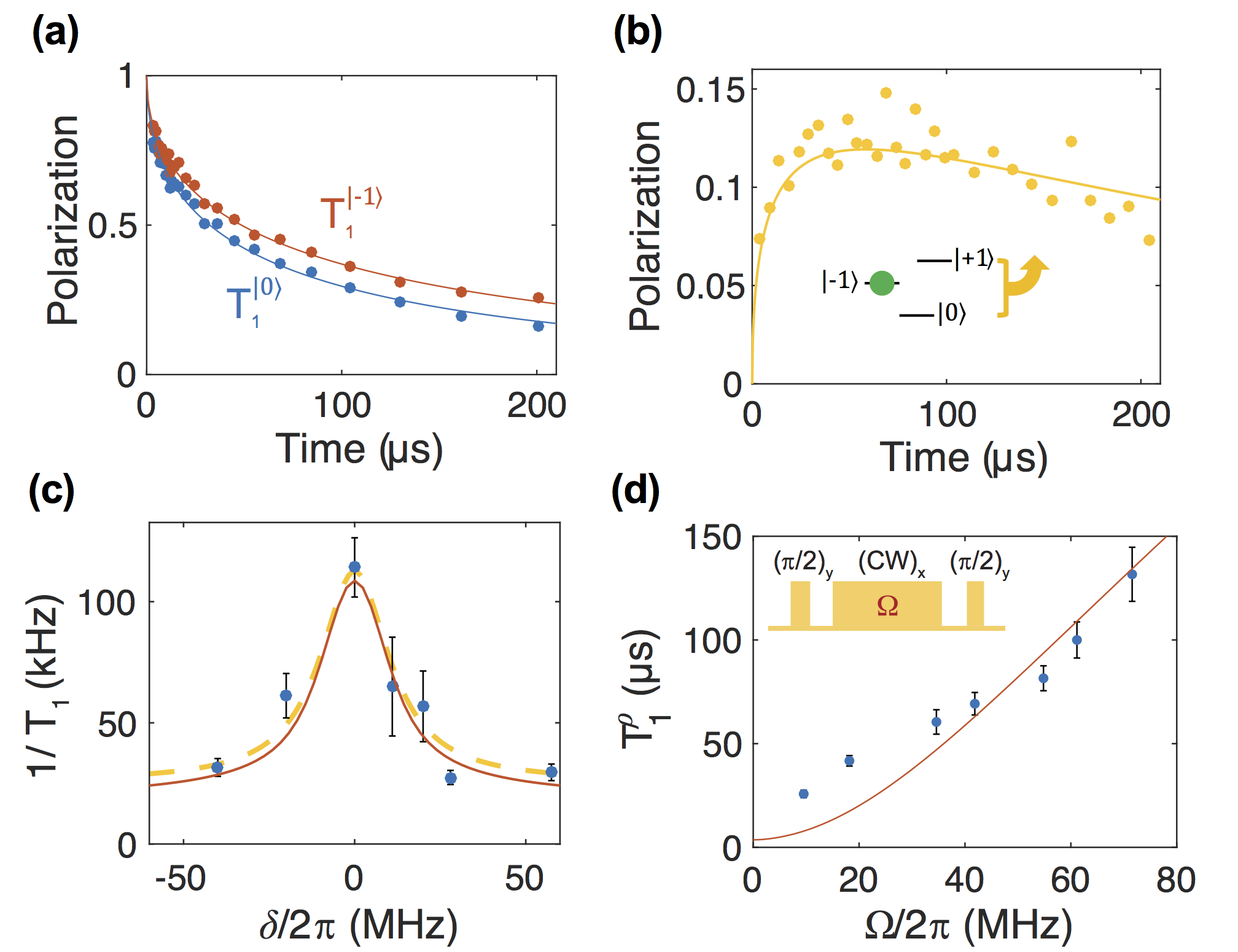}
\caption{\textbf{Depolarization Dynamics.}
\textbf{(a)} NV depolarization timescale probed for different initial states. Solid lines represent stretched-exponential fits with corresponding $T_1$ of 56~$\pm$~2~$\mu$s ($\ket{m_s = 0}$, blue data) and 80~$\pm$~2~$\mu$s ($\ket{m_s = -1}$, red data).
\textbf{(b)} Population difference between $\ket{m_s = 0}$ and $ \ket{m_s = 1}$ as a function of time for initialization into $\ket{m_s = -1}$. Solid line corresponds to a rate equation model of magnetic noise induced spin depolarization~\cite{SOM}.
\textbf{(c)} Measured depolarization rates, $1/T_1$, as a function of the spectral distance, $\delta$, between two groups B and C. 
A Lorentzian fit (dotted orange line) is used to extract the full width at half maximum (FWHM) of $(2\pi)$~25~$\pm$~6~MHz. 
Red line shows the prediction from the spin-fluctuator model at optimized values $n_f=16$~ppm and $\gamma_f=(2\pi)~3.3$~MHz.
\textbf{(d)} Spin-lock lifetime $T_1^\rho$ as a function of driving strength $\Omega$. Red line indicates the prediction from the spin-fluctuator model at an optimized value of $n_f=16$~ppm and $\gamma_f=(2\pi)~3.3$~MHz.} 
\end{figure}

The depolarization dynamics associated with differing spin states is shown in Fig.~2a,b.
For an initial state $\ket{m_s=0}$, we find a decay time of $T_1 \sim 56~\mu$s. For an initial state $\ket{m_s=-1}$, however, we observe an extended lifetime, $T_1 \sim 80~\mu$s, suggesting that the depolarization mechanism is spin-state dependent (Fig.~2a).
This is further confirmed by monitoring the population difference between $\ket{m_s=0}$ and $\ket{m_s=1}$ after initialization into $\ket{m_s=-1}$ (Fig.~2b).
We find that the $\ket{m_s=-1}$ state preferentially decays into the $\ket{m_s =0}$ state, before reaching a maximally mixed state.
Such preferential decay at room temperature is a strong signature that depolarization is induced by an effective magnetic noise~\cite{kolkowitz2015probing}.

Our next set of measurements probes the density dependence of the NV ensemble's relaxation rate.
By tuning the external magnetic field, we can bring two groups of NVs with different orientations of the NV axis into resonance (Fig.~1b).
We monitor the depolarization rate of group B, initialized in $\ket{m_s=0}$, as a function of detuning $\delta$ between group B and C.
As depicted in Fig.~2c, the depolarization rate increases by a factor of $\sim 4$ as the two groups become degenerate, suggesting that interactions among NV centers play an important role in the depolarization mechanism.
Interestingly, the measured width $\Gamma=(2\pi)$~25~$\pm$~6~MHz of this resonant feature significantly exceeds the inhomogeneous linewidth of our sample, $W=(2\pi)~9.3~\pm~0.4$~MHz (extracted from an electron spin resonance measurement) as well as the typical dipolar interaction strength~\cite{kucsko2016critical}.
These results imply that the effective magnetic noise originates from interactions among NV centers with a correlation time $\sim1/\Gamma$.

We further investigate the role of interactions in the depolarization dynamics by performing a spin-locking measurement~\cite{hartmann1962nuclear}.
The spins are initialized  into a superposition state $\ket{+}=(\ket{m_s=0}+\ket{m_s=-1})/\sqrt{2}$ and strong microwave driving is then applied along the axis coinciding with this spin state. 
The driving defines a new dressed-state basis with eigenstates $\ket{+}$ and $\ket{-} = (\ket{m_s=0}-\ket{m_s=-1})/\sqrt{2}$, separated in energy by the Rabi frequency of the microwave field, $\Omega$~\cite{SOM}.
Following time evolution, the population difference between the $\ket{\pm}$ states is measured.

In the context of NMR, spin-locking is known to decouple  nuclear spins  from their environment and to suppress dipolar exchange interactions by a factor of two~\cite{bar2013solid, biercuk2009optimized}.
However, in our case, the combination of spin-locking and the $S=1$ nature of the NV can cause a {\it full} suppression of the flip-flop interactions between the $\ket{\pm}$ states at large $\Omega$~\cite{kucsko2016critical}.
We measure the spin-locking relaxation time, $T_1^\rho$, as a function of $\Omega$ as shown in Fig.~2d. At large $\Omega$, we find that $T_1^\rho$ is extended well beyond the bare lifetime $T_1$.

{\it Spin-Fluctuator Model}---One possible mechanism for the fast, density-dependent depolarization observed above is collectively enhanced spontaneous emission (superradiance)~\cite{dicke1954coherence}.
Indeed, in our system, the average NV separation is well below the wavelength of resonant phonons, potentially enabling multiple spins to couple with a single phonon mode. 
However, due to the large inhomogeneous linewidth $W \gg 1/T_1$, such collective coherences are expected to be strongly suppressed.
The lack of temperature dependence observed in high density samples is also inconsistent with a phonon-limited spin lifetime~\cite{budker2015shortT1}.
Another possible explanation is related to spin-diffusion induced by  dipolar interactions~\cite{cardellino2014effect}.
However, dipolar spin diffusion predominantly affects the boundary of the probed region, and a quantitative estimate suggests a decay which is significantly slower than the observed timescale~\cite{SOM}.

To explain our observations, we now introduce a simple phenomenological model,
in which we assume that a certain fraction of NV centers undergo rapid incoherent depolarization, providing a mechanism for local energy relaxation~\cite{berman2005spin,vugmeister1978spin}.
These short-lived spins (termed fluctuators) can then lead to depolarization of the entire ensemble via dipolar interactions (Fig.~3a).

We now focus on the quantitative analysis of ensemble depolarization arising from the interplay of dipolar interactions, disorder, and dissipative fluctuator dynamics.
Let us assume that fluctuators are randomly positioned in the lattice at density $n_f$ and  depolarize at rate $\gamma_f$ (Fig.~3a).
When  $\gamma_f $ dominates the dipolar interaction strength, each fluctuator can be treated as a localized magnetic noise source with spectral width $2 \gamma_f$ (half width at half maximum).
From Fermi's Golden rule, the depolarization rate of an NV spin induced by a nearby fluctuator is
\begin{align}
    \gamma_s (\vec{r}) \sim \left( \frac{J_0}{r^3}\right)^2 \frac{2\gamma_f}{\delta \omega^2 + (2\gamma_f)^2} 
\end{align}
where $\vec{r}$ is distance between the fluctuator and the spin, 
$J_0 = (2\pi)~52$~MHz$\cdot$nm$^{3}$ is the dipolar interaction strength,
and $\delta \omega$ is the difference in transition frequencies from the inhomogeneous broadening $W$~\cite{SOM}.
For each spin, the effective depolarization rate  is obtained by summing over all fluctuator-induced decay rates: $\gamma^\textrm{eff}_s = \sum_{i \in \textrm{fluctuators}} \gamma_s ( \vec{r}_i)$ (Fig.~3b).
Owing to the random position of fluctuators,  $\gamma^\textrm{eff}_s$ follows a probability distribution 
\begin{align}
\rho(\gamma)=e^{-1/4\gamma T}/\sqrt{4\pi\gamma^3 T}
\end{align}
with the characteristic timescale 
\begin{align}
   \frac{1}{T} = \left(\frac{4\pi n_fJ_0 \eta }{3}\right)^2 \frac{\pi}{\gamma_f}, \label{eq:t1}
\end{align}
where $\eta$ characterizes both the spin exchange matrix element (averaged over all orientations) and the  inhomogeneous broadening~\cite{SOM}.

This model quantitatively captures all  of the  observations in our experiments.
First, resonant dipolar interactions only allow for the exchange of a single unit of spin angular momentum, naturally explaining the  spin-state dependence of the depolarization rate. 
Second, the stretched exponential profile of $P(t)$ arises from integrating over the distribution $\rho(\gamma^\textrm{eff}_s)$;
in particular, while each individual spin undergoes a simple exponential decay, the macroscopic ensemble depolarizes with an averaged profile $P(t)=\int_{0}^\infty \rho(\gamma) e^{-\gamma t} d\gamma = e^{-\sqrt{t/T}}$, precisely matching Eq.~(1) (Fig.~3b).
We emphasize that the functional form of the distribution, $\rho(\gamma)$, results from a combination of dimensionality and the long-range power-law \cite{levitov1990delocalization}; more generally, when the dimension of the system matches the power-law of the interactions, one expects a decay profile,  $P(t)\sim e^{-\sqrt{t/T}}$. 
%
%
 Third, when the two NV groups are tuned into resonance, $\delta=0$, the effective density of fluctuators $n_f$ doubles, thereby enhancing the depolarization rate by a factor of $\sim 4$, consistent with our previous observations.
By computing the effective NV decay rates ($1/T_1$) as a function of $\delta$ and comparing with the experimental data (Fig.~2c), we can extract the density $n_f \sim 16$~ppm and the average decay rate of fluctuators $\gamma_f \sim (2\pi)~3.3$~MHz~\cite{SOM}.
Finally, the extension of the spin lifetime via spin-locking is captured by the suppression of flip-flop interactions~\cite{kucsko2016critical}.
In the ideal case, where the depolarization mechanism results only from resonant exchange, this should lead to a factor of 12 improvement in $T_1^\rho$ as compared to $T_1$~\cite{SOM}.
However, since $T_1^\rho$ is also affected by interactions with NV spins in non-resonant groups, we expect a more modest enhancement in the experiment.
Incorporating both effects, we compare the theory-predicted lifetimes with the experimental data in Fig.~2d, finding reasonable agreement without any additional free parameters.

\begin{figure}[t]
\includegraphics[width=.48\textwidth,left]{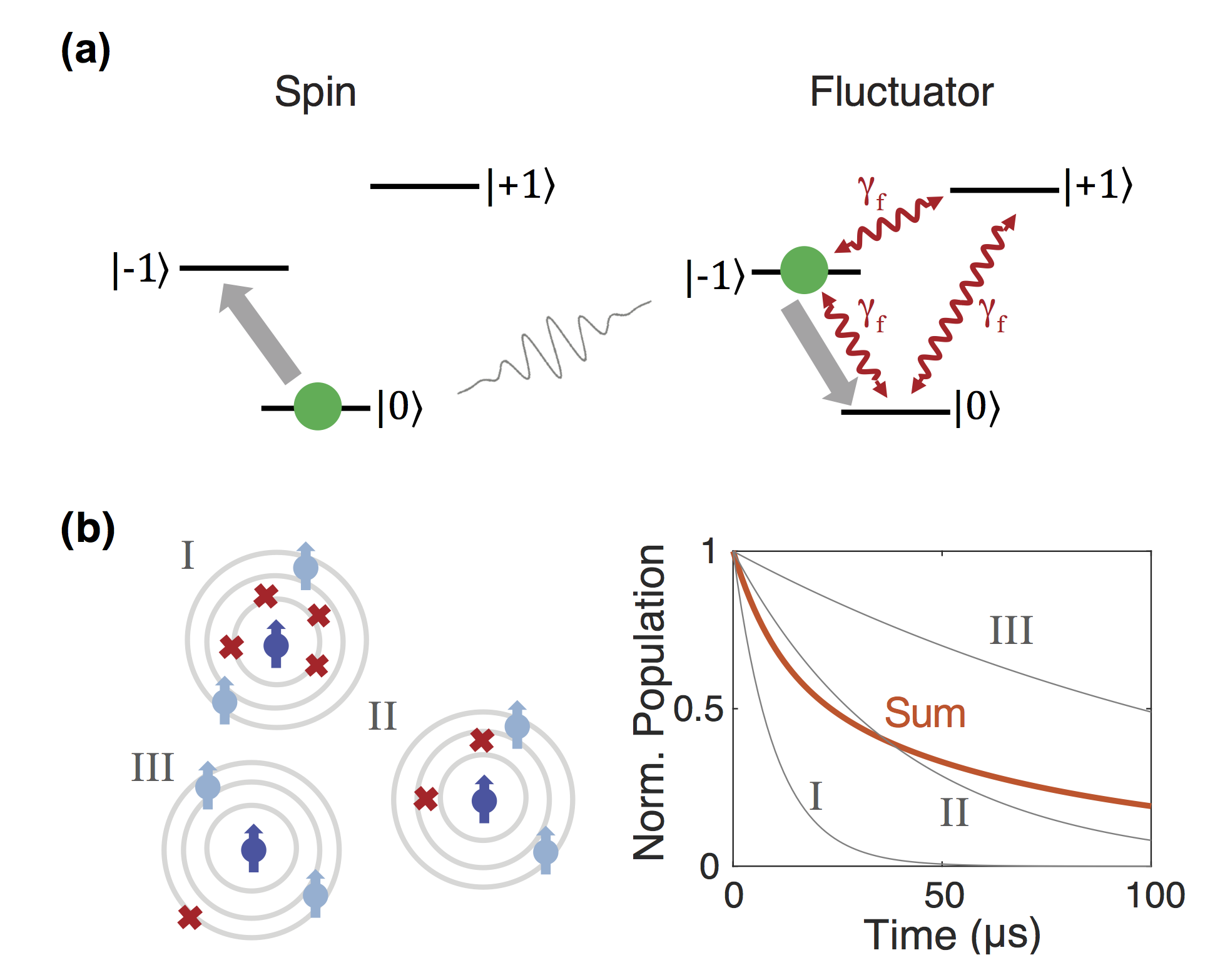}
\caption{\textbf{Spin-Fluctuator Model.} \textbf{(a)} Level diagram of a single spin and fluctuator in two different spin states (green disk). Red arrows indicate fast depolarization channels of a fluctuator.  Solid gray arrows depict spin exchange via dipolar interactions between spin and fluctuator.
\textbf{(b)} Schematic representation of several spins I, II, and III (dark blue) in the ensemble  with different depolarization rates owing to random positions of surrounding fluctuators (red crosses). Ensemble averaging of such depolarization rates gives rise to a stretched exponential with exponent 1/2 (red solid line).}
\end{figure}

{\it Charge Dynamics}---The extracted fluctuator density, $n_f\sim16$~ppm, is a sizable fraction of the 45~ppm of NV spins present in our sample. 
In practice, such fluctuators may arise as a consequence of charge dynamics.
More specifically, electrons may tunnel among a network of closely spaced NV centers, and as the charge state of an NV center changes, its spin state is not necessarily conserved.
We note that such dynamics inevitably occur in high density spin ensembles when impurity wavefunctions overlap and foreshadow the formation of an impurity band~\cite{feher1959electron}.

To probe the existence of such charge hopping, we optically induce a non-equilibrium charge distribution in our bulk diamond sample and monitor the subsequent relaxation back to equilibrium. 
%
%
In the presence of optical illumination at 532~nm, a small fraction of NV centers located at the intensity maximum are excited to the conduction band via a two photon process.
Electrons in the conduction band are delocalized and can recombine with neutral  nitrogen-vacancy defects (NV$^0$) located within a mean free path. 
This charge redistribution can be experimentally measured by scanning a yellow ($\lambda$ = 594~nm) probe laser beam, which selectively excites NV$^-$, relative to the position of the strong ionization beam at $\lambda$ = 532~nm (Fig.~4a,b)~\cite{waldherr2011violation,shields2015efficient}.  
Figure~4c depicts the creation of  a non-uniform charge distribution, with electron depletion at the position of the ionization beam and a surplus in the surrounding regions. 
By monitoring the NV charge state at the origin, after a variable dark interval, we extract a charge recovery time scale of $\sim100$~$\mu$s as illustrated in Figure~4d. 
Interestingly, this recovery occurs in the absence of both optical and thermal excitation, supporting the picture of tunneling-mediated charge diffusion. 
Such fluorescence dynamics have previously been observed in dilute samples on much slower timescales~\cite{iakoubovskii2000photochromism}.
Using a classical diffusion equation, we find a timescale for charge hopping $T_{hop}$ $\sim$10~ns, which is comparable to the independently extracted fluctuator decay time $1/\gamma_f$~\cite{SOM}.
This analysis strongly supports the hypothesis that spin fluctuators are associated with charge hopping between proximal NV centers.

\begin{figure}[t]
\includegraphics[width=.48\textwidth,left]{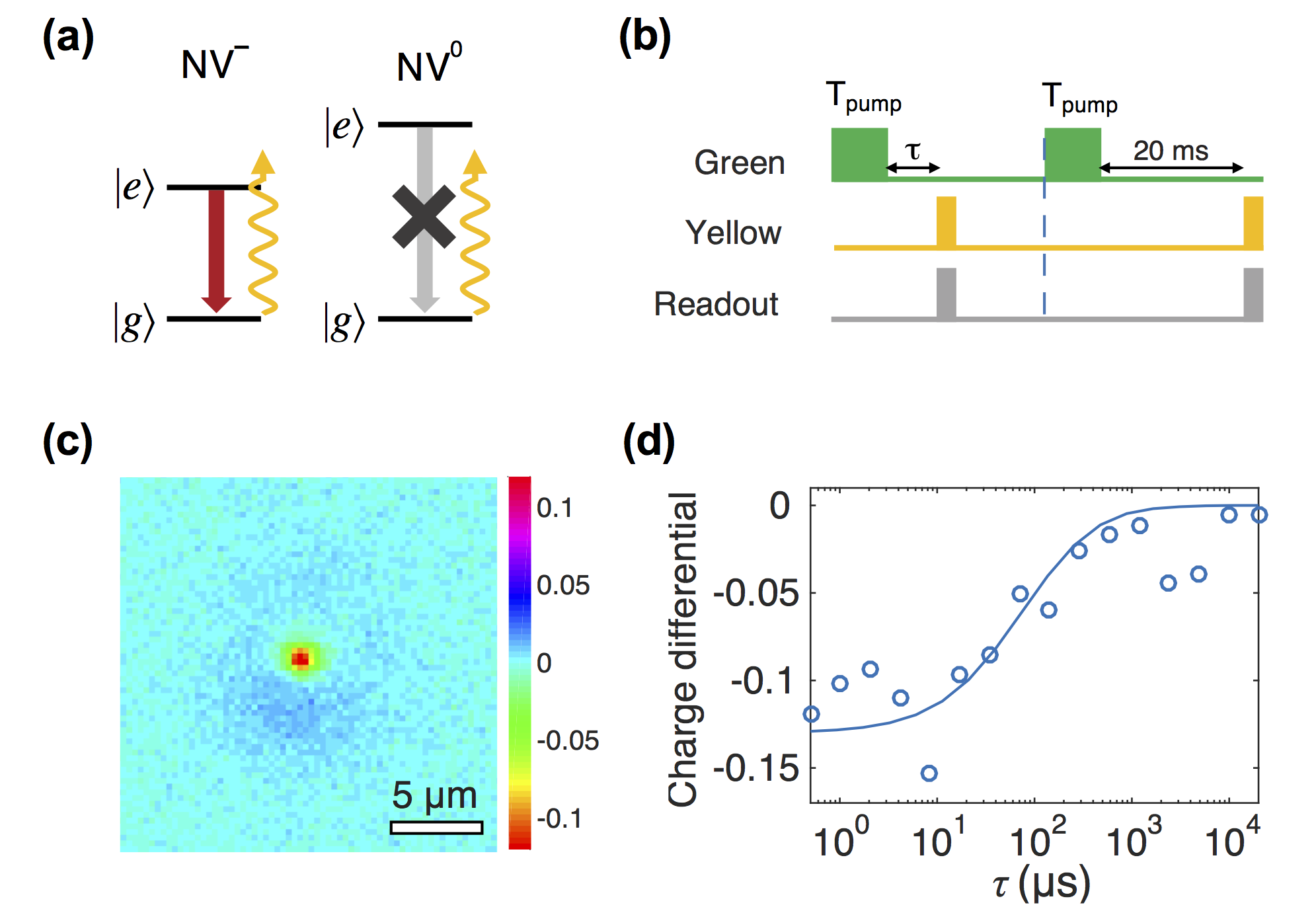}
\caption{\textbf{Charge-State Dynamics} 
\textbf{(a)} Level diagram, showing optical ground state, $\ket{g}$, and excited state, $\ket{e}$, for NV$^{0}$ and NV$^{-}$ under illumination. Yellow laser ($\lambda$ = 594~nm) can off-resonantly excites NV$^{-}$, leading to a strong fluorescence signal. NV$^{0}$, however, remains in its ground state due to a higher transition frequency, allowing optical detection of NV charge states.
\textbf{(b)} Pulse sequence used to measure charge distribution. A green laser is used to create an out-of equilibrium initial state. The resulting charge distribution can be measured via short yellow laser pulses.
\textbf{(c)} Relative charge distribution measured via a yellow scanning laser (27~$\mu$W) after a strong green laser illumination at the center (100~$\mu$W). 
\textbf{(d)} Relaxation of charge distribution at the center over time (open circles) and theoretical fit based on a classical diffusion model (solid line).
}
\end{figure}

{\it Conclusion}---We have investigated the depolarization dynamics in a dense ensemble of interacting NV centers and have proposed a  spin-fluctuator model that quantitatively captures all of the observed dynamics.
Moreover, we suggest a possible microscopic understanding for these fluctuators based on tunneling-mediated charge dynamics. 
We demonstrate that fluctuator-induced depolarization can be mitigated by advanced dynamical decoupling techniques.
In particular, the use of spin-locking allows one to explore coherent dynamics at  long time scale well beyond bare depolarization time~\cite{kucsko2016critical}.
Furthermore, we expect that the depolarization can be controlled by altering the Fermi level via doping~\cite{doi2016pure}.
In such highly doped, disordered systems, experiments of the kind reported here could provide new insights into coupled spin and charge dynamics, complementary to conventional transport measurements.
Our results can be also extended to other solid-state spin systems and provide important guidelines for  understanding the nature of many-body dynamics in  strongly interacting spin ensembles \cite{yao2014many,serbyn2014interferometric}. 

We thank N. P. De Leon, V. Oganesyan, A. Gali, D. Budker, A. Sipahigil, M. Knap and S. Gopalakrishnan for insightful discussions and experimental assistance.
This work was supported in part by CUA, NSSEFF, ARO MURI, Moore Foundation, Miller Institute for Basic Research in Science, Kwanjeong Educational Foundation, Samsung Fellowship, 
NSF PHY-1506284, NSF DMR-1308435, 
Japan Society for the Promotion of Science KAKENHI (No.~26246001), EU (FP7, Horizons 2020, ERC), DFG,  Volkswagenstiftung and BMBF.

\bibliographystyle{apsrev4-1}
\bibliography{depol_bibtex}

\end{document}


\title{Supplementary Information: Depolarization dynamics in a strongly interacting solid state spin ensemble}
\author{J. Choi}
\thanks{These authors contributed equally to this work}
\author{S. Choi}
\thanks{These authors contributed equally to this work}
\author{G. Kucsko}
\thanks{These authors contributed equally to this work}
\author{P. C. Maurer}
\author{B. J. Shields}
\author{H. Sumiya}
\author{S. Onoda}
\author{J. Isoya}
\author{E. Demler}
\author{F. Jelezko}
\author{N. Y. Yao}
\author{M. D. Lukin}

\date{\today}


\maketitle

\tableofcontents

\section{Fluorescence Dynamics}
We utilize a green laser ($\lambda = 532~nm$) to initialize and read out the spin state of NV centers.
Due to a two-photon absorption process this excitation can cause ionization of NV centers, resulting in an out-of-equilibrium charge distribution around the excitation spot.
This charge distribution relaxes even without illumination, as discussed in the main text.
In practice, such charge dynamics can affect the spin-state readout by changing the average fluorescence emission rates (NV$^0$ vs NV$^-$).
In this section, we explain how to avoid this problem.

\begin{figure}[hbt]
\includegraphics[width=1\textwidth]{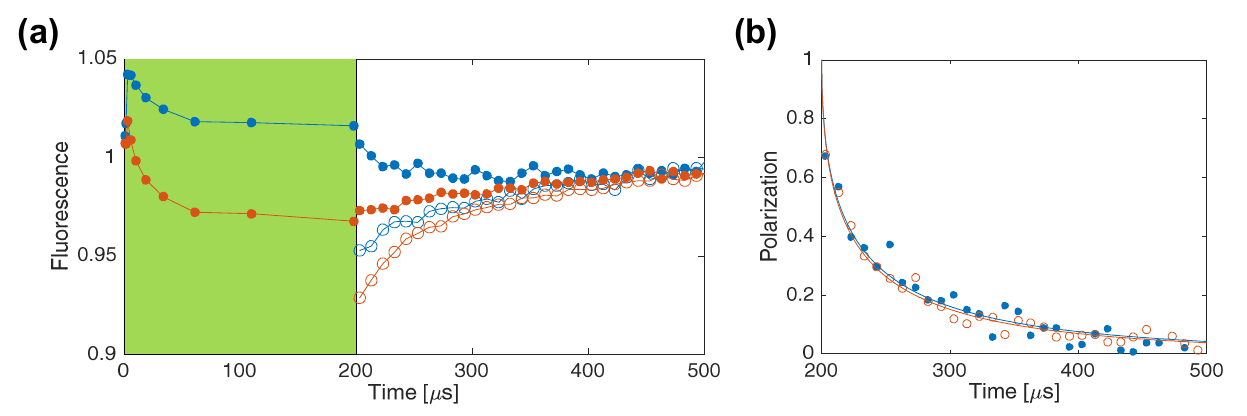} 
\caption{\textbf{Fluorescence Dynamics.} 
\textbf{(a)} Fluorescence emission measurements for two different green laser powers, 10~$\mu$W (solid blue data) 40~$\mu$W (solid red data).
An initial green laser pulse with length $200~\mu$s is used to polarize the spin states of NV centers and reach a charge equilibrium state. After a wait time $t$, another green pulse is used to read out the spin state.
The fluorescence emission rate is measured for both the polarization pulse (green background) and the readout pulse (white background, as a function of $t$).
The emission rate is normalized by an equilibrium value obtained at sufficiently late time ($> 500~\mu$s).
Empty circles correspond to similar measurements with the addition of microwave $\pi$-pulse shortly before readout.
\textbf{(b)} Spin polarization decay as a function of time, extracted from the difference between filled and empty circles from (a) for both excitation powers. 
}  \label{fig:fluorescence}
\end{figure}

The effect of charge dynamics on fluorescence emission rate is illustrated in Fig.~\ref{fig:fluorescence}a.
Under green laser illumination (0-200~$\mu$s) the fluorescence emission rate initially increases as spins are polarized, but then quickly decreases as NV centers get ionized (due to preferential collection from the NV$^-$ phonon sideband).
Following the polarization of spins at varying laser power, we record the florescence after a variable time evolution $t$ with and without an extra microwave $\pi$-pulse at the end of the evolution (empty and full circles Fig.~\ref{fig:fluorescence}a).
We find that the observed fluorescence levels are asymmetric and for high power increase as a function of time.
This fluorescence increase is caused by previously ionized NV centers, relaxing back to equilibrium.
This effect is particularly dominant at high green excitation power, where ionization at the focal spot is increased.
However, as demonstrated in Fig~\ref{fig:fluorescence}b, the difference in fluorescence with and without the extra $\pi$-pulse is independent from the applied green laser power. 
This result implies that one can use such a differential readout to mediate the contribution of charge dynamics and reliably extract the spin depolarization time scales.
Additionally, the use of a low power green laser ($\sim$10~$\mu$W) can help reduce the effects of charge dynamics during the experiment.
Note, that a similar technique has previously been used to robustly measure spin dynamics in high density NV samples~\cite{budker2015shortT1}.

\section{Estimation of Dipolar Spin Diffusion Timescale}
One potential mechanism of spin-density dependent depolarization is that polarized spins diffuse out of the probing volume via dipolar interactions.
Here, we estimate the timescale of such spin diffusion using a classical diffusion equation.
In our experiments, the probing volume is determined by the confocal excitation spot size $w\sim 200~$nm. Note that this estimate assumes a bulk diamond excitation, approximated by a 2D model, due to the large extent of the excitation spot in the direction normal to the diamond surface. The effective diffusion coefficient $D$ can then be estimated from the average spacing among NV centers $a\sim5~$nm and the typical flip-flop time $\tau \sim 9.5~\mu$s (calculated from $J = (2\pi)~105$~kHz), $D \approx a^2 / \tau \sim 2.6~\textrm{nm}^2/\mu\textrm{s}$.
Assuming that the spin polarization initially follows a Gaussian distribution with spatial width $w$, the classical diffusion equation predicts the polarization profile at later time $t$
\begin{align}
P(t,\vec{r}) = \frac{e^{- \frac{r^2}{2 (w^2+Dt)}}}{2(w^2+Dt)\pi},
\end{align}
where one finds that the spin polarization density at $r=0$ is reduced by a factor of 2 at time $t \sim  w^2/D \sim 15~$ms. This timescale is more than two orders of magnitude slower than experimentally measured depolarization times.
We note that our estimation ignores the effect of inhomogeneous distribution of NV transition frequencies arising from the presence of other magnetic impurities in diamond; such disorder in transition energies further suppresses resonant spin flip-flop dynamics.
For this reason, we rule out spin diffusion as the sole mechanism of ensemble depolarization.

\section{Rate Equation Model: Spin-state Dependent Depolarization}

\begin{figure}[t]
\includegraphics[width=0.9\textwidth]{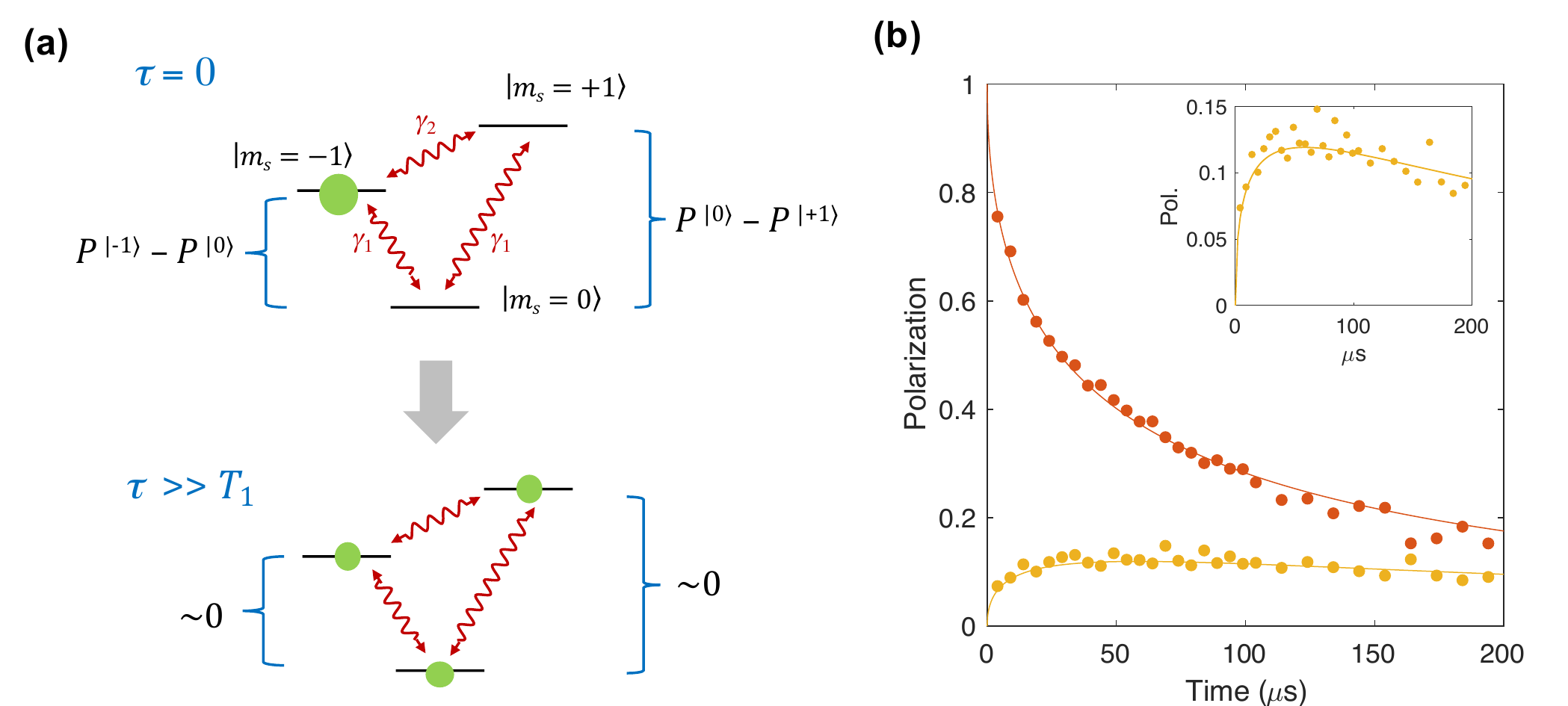} 
\caption{\textbf{Magnetic Noise Model.} 
\textbf{(a)} Schematic level diagram depicting population distribution at $\tau=0$ (after initialization into $|-1\rangle$) and at $\tau>>T_1$. Possible relaxation channels with rates $\gamma_1$ and $\gamma_2$ are indicated as red arrows.
\textbf{(b)} Population difference over time between $|-1\rangle$ and $|0\rangle$ (red data) as well as between $|+1\rangle$ and $|0\rangle$ (orange data) after initialization into $|-1\rangle$. Solid lines indicate the results of the rate equation model, fit to the data. Inset shows a zoom-in of the $|+1\rangle$ - $|0\rangle$ data for better visibility.
}  \label{fig:rateeq}
\end{figure}

In order to estimate the decay rates of two independent depolarization channels $\gamma_1$ and $\gamma_2$ (Fig.~1a, main text), we analyze the population changes in $|m_s=0\rangle$ and $|m_s=+1\rangle$ after initialization into $\ket{m_s=-1}$ (Fig.~\ref{fig:rateeq}b).
For this we employ the following simple rate equation model
\[
\frac{d}{dt} \begin{bmatrix}
   P^{\ket{-1}} \\
   P^{\ket{0}} \\
   P^{\ket{+1}}
\end{bmatrix}
=
\begin{bmatrix}
    -\gamma_1-\gamma_2 & \gamma_1 & \gamma_2  \\
    \gamma_1 & -2\gamma_1 & \gamma_1   \\
    \gamma_2 & \gamma_1 & -\gamma_1-\gamma_2   
\end{bmatrix} 
\begin{bmatrix}
   P^{\ket{-1}} \\
   P^{\ket{0}} \\
   P^{\ket{+1}} 
\end{bmatrix},
\]
where $P^{\ket{-1}}, P^{\ket{0}}$, and $P^{\ket{+1}}$ are the normalized populations in each spin states. When spins are initialized into $\ket{m_s = -1}$ at $t = 0$, the solution of the rate equation model predicts 
\begin{align}
P^{\ket{-1}}(t) - P^{\ket{0}}(t) &=  \frac{1}{2} e^{-(\gamma_1 + 2\gamma_2) t} + \frac{1}{2}e^{-3\gamma_1 t}    \label{eq:pop1_exp}  \\
P^{\ket{0}}(t) - P^{\ket{+1}}(t) &=  \frac{1}{2} e^{-(\gamma_1 + 2\gamma_2) t} - \frac{1}{2}e^{-3\gamma_1 t}.  \label{eq:pop2_exp}
\end{align}
In our experiments, however, the decay rates $\gamma_1$ and $\gamma_2$ are random variables, giving rise to stretched exponential profiles.
Consequently, the measured population differences become
\begin{align}
P^{\ket{-1}}(t) - P^{\ket{0}}(t) &=   \frac{1}{2} e^{-\sqrt{(\gamma_1 + 2\gamma_2)t}} + \frac{1}{2}e^{-\sqrt{3\gamma_1 t}}  \label{eq:pop1}  \\
P^{\ket{0}}(t) - P^{\ket{+1}}(t) &=  \frac{1}{2} e^{-\sqrt{(\gamma_1 + 2\gamma_2)t}} - \frac{1}{2}e^{-\sqrt{3\gamma_1 t}} .  \label{eq:pop2}
\end{align} 
Fitting Eq. (\ref{eq:pop1}) and (\ref{eq:pop2}) to experimental data, we extract ${\gamma}_1 = 10.6 \pm 0.6$ kHz and ${\gamma}_2 = 1.1 \pm 0.7 $ kHz.
As shown in Fig.~\ref{fig:rateeq}b, this simple theory prediction and the experiment data are in excellent agreement.
In particular, we notice that $\gamma_1 \gg \gamma_2$, implying that the spin state decay is induced by a local magnetic noise which changes only one unit of magnetization at a time, $\Delta m_s = \pm 1$.

\section{Detailed Analysis of Spin-Fluctuator Model}
In this section, we provide a detailed analysis of our fluctuator model.
As described in the main text, we assume that a fraction of NV centers, with density $n_f$, undergo rapid depolarization at rate $\gamma_f$. Their positions and orientations are randomly distributed.
These fluctuators interact with normal spins via dipolar interactions, inducing depolarization. 
Here, we derive four characteristic features of the observed depolarization dynamics presented in the main text: (a) spin-state dependent polarization rates and preferential decay, (b) stretched exponential decay of ensemble polarization, (c) the resonant feature of depolarization rates as two groups of NV centers become degenerate, and (d) the extension of spin lifetime via spin-locking.
Note that the quantitative analysis in (c) allows us to extract the values of $\gamma_f$ and $n_f$ from the experimental data presented in the main text (Fig. 2c main text).
We use those extracted parameters to predict the spin lifetime $T_1^\rho$ under spin-locking, which shows reasonable agreement with the data (Fig.~2d main text).
This section is organized in the following order.
First, we compute the effective depolarization rate of a normal spin induced by dipolar interaction with a single fluctuator. We will show that such a mechanism results in spin-state dependent $T_1$ and preferential decay.
Second, we show that interactions with multiple fluctuators located at random positions results in a distribution of effective decay rates $\rho(\gamma)$.
We will explicitly compute the ensemble depolarization $P(t) = \int_0^\infty \rho(\gamma) e^{-\gamma t} d\gamma = e^{-\sqrt{t/T_1}}$ and provide an analytic expression for $T_1$ in terms of microscopic parameters.
Finally, we will apply these results to various scenarios to predict the resonant features in Fig.~2c of the main text and the extension of spin lifetime in Fig.~2d of the main text.

\subsubsection*{Single Spin Interacting with a Single Fluctuator}
\begin{figure}[t]
\includegraphics[width=0.5\textwidth]{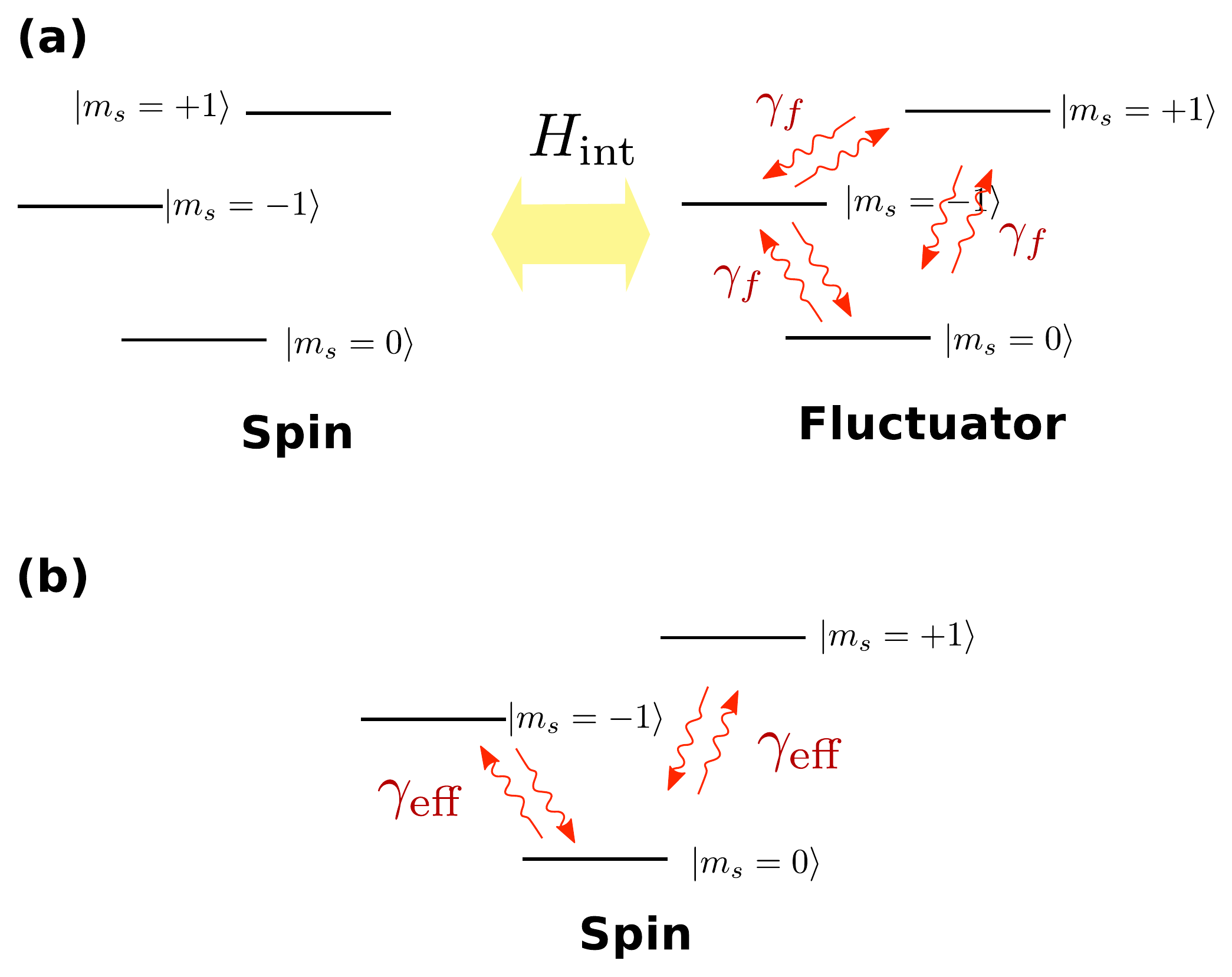} 
\caption{\textbf{Spin-Fluctuator interaction} 
\textbf{(a)} Level diagram showing a single spin interacting with a single fluctuator.
The fluctuator undergoes rapid incoherent depolarization at a rate $\gamma_f$. 
\textbf{(b)} Effective incoherent dynamics of a spin due to fluctuator interactions. 
}  \label{fig:level_diagram}
\end{figure}

We begin our analysis by considering a system of a single spin and a single fluctuator that interact via  Hamiltonian $H_\textrm{int}$.
The total Hamiltonian of such a system is given as
\begin{align}
    H = H_1 + H_2 + H_\textrm{int}
\end{align}
where $H_1$ and $H_2$ are single particle Hamiltonians for a normal spin and a fluctuator, respectively. The details of $H_1$ and $H_2$ varies over different experiments. For example, in typical $T_1$ measurements where an external magnetic field is aligned along the quantization axis of NV centers, the single particle Hamiltonians are diagonal in the natural spin basis $H_{1/2} = \sum_{m_s\in \{0,\pm1\}} \omega_{m_s} \ket{m_s}\bra{m_s}$, where the energy eigenvalues $\omega_{m_s}$ in the rotating frame are random number of order  $W\sim (2\pi)~9$~MHz, owing to inhomogeneous broadening of the system.
When strong spin-locking with Rabi frequency $\Omega$ is applied between $\ket{m_s=0}$ and $\ket{m_s=-1}$ transition, a dressed state basis is preferred, where
 $H_{1/2} = \pm (\Omega/2) \ket{\pm}\bra{\pm} + \omega_0 \ket{m_s=0} \bra{m_s=0}$ with $\ket{\pm} = (\ket{m_s=0}\pm\ket{m_s=-1})/\sqrt{2}$ as defined in the main text.
 Later, we will also consider a situation where $H_1$ and $H_2$ are diagonal in the different basis. Such a case arises when the normal spin and the fluctuator are oriented in different directions. Here, for simplicity, we assume a generic eigenbasis $\{\ket{i}\}$ and $\{\ket{\alpha}\}$ for $H_1$ and $H_2$, so that 
\begin{align}
    H_1 = \sum_{i\in\{1, 2, 3\}} \omega_i \ket{i}\bra{i} \;\; \textrm{and} \;\;
    H_2 = \sum_{\alpha\in\{a,b,c\}} \omega_\alpha \ket{\alpha}\bra{\alpha},
\end{align}
where $\omega_i$ and $\omega_\alpha$ are corresponding energies in the rotating frame.
In addition to coherent dynamics, the fluctuator undergoes rapid incoherent dynamics. Hence, the dynamics of the system are governed by a quantum master equation
\begin{align}
    \dot{\rho} &= -i [H, \rho ] + L[\rho] \\
    L[\rho] &= \sum_k L_k \rho L_k^\dagger - \frac{1}{2} \left( L_k^\dagger L_k \rho + \rho L_k^\dagger L_k\right),
\end{align}
where $\rho$ is the density matrix of the system and $L_k$ are decay operators. In our model, we consider six decay processes as illustrated in Fig.~\ref{fig:level_diagram}(a) with identical decay rates $\gamma_f$, i.e., $L_k = \sqrt{\gamma_f} \ket{m_s=\alpha} \bra{m_s=\beta}$ with $(\alpha,\beta) \in \{ (+1,-1),(-1,+1),(+1,0),(0,+1),(-1,0),(0,-1)\}$.

In order to derive the effective master equation for a spin, we use the Born-Markov approximation together with secular approximations \cite{gardiner2004quantum}.
In such a description, the quantum state of the system is approximated by 
$\rho = \rho_1 \otimes \rho_\textrm{thm}$, where $\rho_1$ is the reduced density matrix of the normal spin and $\rho_\textrm{thm}$
is the equilibrium state of the fluctuator, such that $L[\rho_1\otimes \rho_\textrm{rhm}]=0$.
In our model, the equilibrium state is a maximally mixed state $\rho_\textrm{thm} = \frac{1}{3} I$.
These approximations are well justified due to the hierarchy of the coupling strengths $J_{ij} \ll |\omega_{ij} |  \ll \gamma_f$, where $\omega_{ij} =\omega_i -\omega_j$ are the energy difference between eigenstates and $J_{ij}$ are typical dipolar interaction strengths.
Under these approximations, the effective dynamics of a spin become  
\begin{align}
\dot{\rho_1} = -i [H_1, \rho_1] + L^\textrm{eff}[\rho_1],
\end{align}
where the first and second terms describe the coherent dynamics by the Hamiltonian $H_1$ and the induced dynamics by interactions with the fluctuator, respectively. Following standard procedures \cite{gardiner2004quantum}, we obtain
\begin{align}
L^\textrm{eff}[\rho ]  
\approx& 
\sum_{i,j} \sum_{\alpha \beta \gamma \delta }
 C^{ij}_{\alpha \beta}C^{ji}_{\gamma \delta} S^{\alpha \beta \gamma \delta} (\omega_{ij}) 
 \Big[\ketbra{j}{i} \rho \ketbra{i}{j} - \ketbra{i}{i} \rho  \Big]+ h.c.
\label{eqn:003}\\
 & + \sum_{i\neq k} \sum_{\alpha \beta \gamma \delta} 
 C^{ii}_{\alpha \beta}
C^{kk}_{\gamma \delta }  
 S^{\alpha \beta \gamma \delta} (\omega_{ij}) 
\Big[
  \ketbra{k}{k} \rho \ketbra{i}{i} - \ketbra{i}{i}\ketbra{k}{k} \rho 
\Big] + h.c.
\label{eqn:005}
\end{align}
where $C^{ij}_{\alpha \beta}$ is a matrix element of the interaction defined as $C^{ij}_{\alpha \beta} \equiv \bra{i\alpha} H_\textrm{int} \ket{j \beta}$ with  $i, j \in \{1,2, 3\}$ and  $\alpha, \beta \in \{a, b, c\}$, and $S^{\alpha \beta \gamma \delta}$ is the spectral response function of the fluctuator defined as
\begin{align}
S^{\alpha \beta \gamma \delta} (\omega) = \int_0^\infty e^{i \omega \tau} \bra{\beta}  e^{ \tau \mathcal{L}_2} \Big[ \ket{\gamma} \bra{\delta} \rho_\textrm{thm}\Big] \ket{\alpha}.
\end{align}
where the superoperator $\mathcal{L}_2[\cdot] = -i[H_2, \cdot] + L[\cdot]$ describes the time evolution of the fluctuator. For the decay channels illustrated in Fig.~\ref{fig:level_diagram}, the spectral response function $S^{\alpha \beta \gamma \delta}$ can be simplified by
\begin{align}
S^{\alpha \beta \gamma \delta} (\omega) =  \delta_{\beta \gamma } \delta_{\alpha \delta} S^{\alpha \beta} (\omega) \equiv \delta_{\beta \gamma } \delta_{\alpha \delta} 
\left\{
\begin{array}{cc}
\frac{1}{3} \cdot\frac{ 1 }{i (\omega +\omega_{\alpha \beta})- 2\gamma_f} & \textrm{ for } \alpha \neq \beta\\
\frac{1}{9} \cdot \left( \frac{1}{i\omega} + \frac{2}{ i\omega - 3\gamma_f}\right) & \textrm{ for } \alpha = \beta
\end{array}
\right.
.
\end{align}
Since $H_\textrm{int}$ is Hermitian, we can use the relation $C^{ij}_{\alpha \beta} = (C^{ji}_{\beta \alpha})^*$ to obtain 
\begin{align}
L^\textrm{eff}[\rho ]  
\approx& 
\sum_{i,j} \sum_{\alpha \beta }
 |C^{ij}_{\alpha \beta}|^2 S^{\alpha \beta } (\omega_{ij}) 
 \Big[\ketbra{j}{i} \rho \ketbra{i}{j} - \ketbra{i}{i} \rho  \Big]+ h.c.
\label{eqn:006}\\
 & + \sum_{i\neq k} \sum_{\alpha \beta } 
 C^{ii}_{\alpha \beta}
C^{kk}_{\beta \alpha }  
 S^{\alpha \beta } (0) 
\Big[
  \ketbra{k}{k} \rho \ketbra{i}{i} - \ketbra{i}{i}\ketbra{k}{k} \rho 
\Big] + h.c.
\label{eqn:009}
\end{align}
Finally, introducing 
\begin{align}
\label{eqn:generic_decay_rate}
    \Gamma_{ij} = 2 
    \sum_{\alpha \beta}
    |C^{ij}_{\alpha \beta}|^2
    \textrm{Re}\Big[S^{\alpha \beta} (\omega_{ij})\Big]\;\;\; \& \;\;\;
    \Delta_{ij} = 2 
    \sum_{\alpha \beta}
     |C^{ij}_{\alpha \beta}|^2 
    \textrm{Im}\Big[S^{\alpha \beta} (\omega_{ij})\Big],
\end{align}
the effective superoperator simply becomes
\begin{align}
    L^\textrm{eff} [\rho]
    =& \sum_{ij} \Gamma_{ij} 
    \Big [
        \ketbra{j}{i} \rho \ketbra{i}{j}
        - \frac{1}{2}
            \left( \ketbra{i}{i} \rho 
                + \rho \ketbra{j}{j} \right)
    \Big ]\\
    & -i \sum_{ij} \Delta_{ij} \big[ \ketbra{i}{i}, \rho \big] \\
    & + \sum_{i\neq k} \sum_{\alpha \beta} 
 C^{ii}_{\alpha \beta}
C^{kk}_{\beta \alpha }  
 S^{\alpha \beta } (0) 
\Big[
  \ketbra{k}{k} \rho \ketbra{i}{i}
\Big] + h.c.
\end{align}
Here, we clearly see three types of terms: (i) depolarization or dephasing at rates $\Gamma_{ij}$, (ii) energy corrections $\Delta_{ij}$ to coherent dynamics, and (iii) additional dephasing from diagonal interactions. Here, we are most interested in the depolarization processes $\Gamma_{ij}$ under various conditions.
In particular, we note that the dipolar interaction under secular approximation is given by 
\begin{align}
\label{eqn:secular_dipole_interaction}
    H_{dd}\approx
    - J_0 /r^3 
    \big[ 
            \left( g + i h\right) \left( \ketbra{+1,0}{0,+1}
        + \ketbra{0,-1}{-1,0}\right)
        + h.c.
        + q S^z_i S^z_j
    \big],
\end{align}
where $J_0 = (2\pi)~52~\text{MHz}\cdot\text{nm}^3$ is the dipolar interaction strength, $r$ is the distance, and $g$, $h$, and $q$ are coefficients of order unity that depend on the relative orientation of the spin and fluctuator:
\begin{align}
    g =& \frac{1}{2}
    \Big[ 
            3\left(\hat{r}\cdot \hat{x}_s\right)\left(\hat{r}\cdot \hat{x}_f\right)
            -
            \hat{x}_s \cdot\hat{x}_f
            +
             3\left(\hat{r}\cdot \hat{y}_s\right)\left(\hat{r}\cdot \hat{y}_f\right)
            -
            \hat{y}_s \cdot\hat{y}_f
    \Big]\\
     h =& \frac{1}{2}
        \Big[
            3\left(\hat{r}\cdot \hat{x}_s\right)\left(\hat{r}\cdot \hat{y}_f\right)
            -
            \hat{x}_s \cdot\hat{y}_f
        -
            3\left(\hat{r}\cdot \hat{y}_s\right)\left(\hat{r}\cdot \hat{x}_f\right)
            +
            \hat{y}_s \cdot\hat{x}_f
        \Big]\\
        q = &             3\left(\hat{r}\cdot \hat{z}_s\right)\left(\hat{r}\cdot \hat{z}_f\right)
            -
            \hat{z}_s \cdot\hat{z}_f
\end{align}
with unit vectors $(\hat{x}_a,\hat{y}_a,\hat{z}_a)$ characterizing the quantization axis of the spin ($a=s$) or fluctuator ($a=f$) \cite{kucsko2016critical}.
Importantly, this interaction does not contain any transitions between $\ket{m_s=+1}$ to $\ket{m_s=-1}$, resulting in vanishing decay rates between the two states, i.e. $\Gamma_{+1,-1} = \Gamma_{-1,+1} = 0$.
Consequently, the interaction-induced dynamics of a spin can be modeled as in Fig.~\ref{fig:level_diagram}, which exhibit spin-state dependent depolarization rates as well as preferential decays described in the previous section.
More specifically, the induced decay rate is 
\begin{align}
\gamma = \frac{J_0^2}{r^6} \left(  |g^2| + |h|^2\right)  \frac{2}{3} \frac{ 2\gamma_f }{(\delta\omega )^2 +4\gamma_f^2} \equiv \frac{J_0^2}{r^6} \frac{s^2}{\gamma_f},
\end{align}
where $\delta\omega$ is the energy difference (due to inhomogeneous broadening) between the spin and the fluctuator, and $s$ is a dimensionless number of order unity that characterizes the orientation dependent coefficients of the dipolar interaction as well as spectral responses
\begin{align}
s^2=   \frac{2}{3}  (|g|^2 + |h|^2)\frac{2\gamma_f^2}{ \delta\omega^2 + 4 \gamma_f^2}.
\end{align}

\subsubsection*{Derivation of the Stretched Exponential Decay}
In an ensemble, the net decay rate $\gamma_s^\textrm{eff}$ of a spin is given as the sum of rates induced by multiple nearby fluctuators. Consequently, the decay rates vary from one spin to another, and the ensemble polarization decays as sum of multiple simple exponentials, whose temporal profile depends on the probability distribution of effective decay rates $\rho(\gamma)$.
Here, we compute this distribution and show that the ensemble polarization decays as a stretched exponential
\begin{align}
    \label{eqn:stretched_exp}
    P(t) = e^{-\sqrt{t/T_1}}.
\end{align}
We will see that the exponent, $1/2$, arises as a consequence of incoherent dipole-dipole interaction in 3D. We start with a single spin located at the origin $\vec{r}=0$ and consider its effective depolarization rate induced by fluctuators at $\{ \vec{r}_1, \vec{r}_2, \dots, \vec{r}_N\}$. The polarization decays as a simple exponential with the rate given by $\gamma^\textrm{eff}_s = \sum_i \gamma_i$, where $\gamma_i$ is the decay rate induced by fluctuator $i$. From the previous section, we have shown that each $\gamma_i$ can be written as $\gamma_i = \frac{J_0^2}{r_i^6} \frac{ s_i^2}{\gamma_f}$.
The probability distribution $\rho(\gamma_s^\textrm{eff})$ is obtained by averaging $\delta \left( \sum_i \gamma_i - \gamma_s^\textrm{eff}\right) $ over all possible configurations of fluctuators: different number $N$, positions, and orientations
\begin{align}
\rho(\gamma_s^\textrm{eff}) =  \int d \{ \vec{r}_i \}  \; \textrm{Prob}( \{ \vec{r}_i \} ) 
    \;
   \delta \left( \sum_i \gamma_i - \gamma_s^\textrm{eff}\right).
\end{align}
When the positions of fluctuators are homogeneously distributed with density $n_f$, we can analytically compute $\rho(\gamma_s^\textrm{eff})$:
\begin{align}
   \rho(\gamma_s^\textrm{eff})  =& \sum_N \int_\mathcal{D} dr_1 \dots dr_N 
    \left(
        \prod_{i=1}^N e^{- \frac{4 \pi}{3} n_f (r_i^3 - r_{i-1}^3)} 4 \pi n_f r_i^2 
    \right)
    e^{- \frac{4 \pi}{3} n_f (R^3 - r_{N}^3)}
    \left\langle    
    \delta \left( \sum_i \gamma_i - \gamma_s^\textrm{eff}\right)\right\rangle_s \\
    =& \int_{-\infty}^\infty dz \frac{e^{i \gamma_s^\textrm{eff} z}}{\sqrt{2\pi}}  \sum_N \int_\mathcal{D} dr_1 \dots dr_N 
    \left(
        \prod_{i=1}^N e^{- \frac{4 \pi}{3} n_f (r_i^3 - r_{i-1}^3)} 4 \pi n_f r_i^2 
        \left\langle        
        e^{- i \gamma_i z}
        \right\rangle_s
    \right)
    e^{- \frac{4 \pi}{3} n_f (R^3 - r_{N}^3)}    
\end{align}
where the domain of the integral is $\mathcal{D} = r_0 \leq r_1 \leq r_2 \dots \leq r_N \leq R$ with the shortest (largest) distance cut-off $r_0$ $(R)$, $z$ is a dummy variable introduced for $\delta(x) = \int_{-\infty}^\infty dz e^{ixz}/\sqrt{2\pi} $, and $\left \langle \cdot \right\rangle_s$ represents the averaging of all possible orientations of a fluctuator with fixed distance. We denote this distribution with $\textrm{Prob}(s)$. Now, we see that
\begin{align}
    \rho(\gamma_s^\textrm{eff})  =&  \int_{-\infty}^\infty dz \frac{e^{i \gamma_s^\textrm{eff} z}}{\sqrt{2\pi}}    e^{-\frac{4\pi}{3} n_f (R^3 - r_0^3)} \; \sum_N \frac{1}{N!} 
    \left[ 
       \int_{r_0}^R     4 \pi n_f r^2 dr 
       \int \textrm{Prob}(s) \; ds \; e^{- i\frac{J_0^2}{\gamma_f r_i^6} s^2 z}         
    \right]^N\\
    =& \int_{-\infty}^\infty dz \frac{e^{i \gamma_s^\textrm{eff} z}}{\sqrt{2\pi}} 
    e^{-\frac{4\pi}{3} n_f (R^3 - r_0^3)} \; \sum_N \frac{1}{N!} 
    \left[ 
       \int_{u_0}^U    \frac{ 4 \pi n_f }{3}  du 
       \int \textrm{Prob}(s) \; ds \; \sqrt{(J_0^2 s^2/\gamma_f)|z|} e^{-i \frac{\textrm{sgn}(z)}{u^2}}         
    \right]^N
\end{align}
where we introduced $u = r^3/\sqrt{(J_0^2 s^2/\gamma_f)|z|}$ and similarly $u_0$ and $U$ for $r=r_0$ and $r=R$.
Here, the integration over $u$ can be done analytically. 
Note that we are interested in the limit of large $R$ and small $r_0$, which corresponds to $u_0 \ll 1$ (behavior at long enough time) and $U \gg 1$ (before the boundary effect becomes relevant).
\begin{align}
\int_{u_0}^U du e^{-i \; \textrm{sgn}(z)/u^2} \approx U -  (1 + \textrm{sgn}(z) i ) \sqrt{\pi/2}.
\end{align}
Finally, we obtain
\begin{align}
     \rho(\gamma)  \approx&  \int_{-\infty}^\infty dz \frac{e^{i \gamma z}}{\sqrt{2\pi}}
     e^{-\frac{4\pi n_f}{3} \sqrt{\pi/2}(1 + \textrm{sgn}(z) i ) \int \textrm{Prob}(s) \sqrt{(J_0^2 s^2/\gamma_f) |z|} ds}\\
     =&\int_{-\infty}^\infty dz \frac{e^{i \gamma z}}{\sqrt{2\pi}} e^{(iz/T)^{1/2}} 
     = \frac{e^{-1/(4\gamma T)}
}{\sqrt{4\pi \gamma^3 T}} \; \end{align}
where we introduced the time scale 
\begin{align}
    \frac{1}{T}  \equiv& \left(\frac{4\pi n_fJ_0  \eta }{3}\right)^2 \frac{\pi}{\gamma_f} 
    \label{eqn:Tstr}
\end{align}
with the orientation averaged  $ \eta \equiv \int \textrm{Prob}(s) \; s\; ds$.
The ensemble depolarization profile $P(t)$ can be computed from $\rho(\gamma_s^\textrm{eff})$:
\begin{align}
P(t) = \int_0^\infty \rho(\gamma) e^{-\gamma t} d\gamma = e^{-\sqrt{t/T}}.
\end{align}

\subsubsection*{Enhanced Depolarization of Two Degenerate Groups of NV Centers}
When all four groups of NV centers with different quantization axes are spectrally separated, e.g. in Fig.~1b upper curve in the main text,
the spin exchange interactions between NV centers in distinct groups are strongly suppressed due to a large energy mismatch.
In such case, the depolarization dynamics of a spin are dominated by interactions with fluctuators within the same group.
When two groups of NV centers are brought onto resonance, e.g. Fig.~1b lower curve in the main text, the inter-group dipolar interactions cannot be neglected, resulting in an enhanced effective depolarization rate.
This effect can be quantitatively analyzed by modifying the probability distribution
\begin{align}
\textrm{Prob}(s) = \frac{1}{4}\; \textrm{Prob}^\textrm{same}(s) + \frac{1}{4} \;\textrm{Prob}^\textrm{diff}(s) + \frac{1}{2} \cdot 0,
\end{align}
where $\textrm{Prob}^\textrm{same}(s)$ and $\textrm{Prob}^\textrm{diff}(s)$ correspond to the probability distributions of $s$ for dipolar interactions within a group and between two near-resonant groups, respectively. The other two groups with probability $1/2$ do not induce resonant depolarization.
Crucially, the latter distribution depends on the spectral distance $\delta$ because $s$ is a function of energy mismatch between a spin and a fluctuator:
\begin{align}
s^2=   \frac{2}{3}  (|g|^2 + |h|^2)\frac{2\gamma_f^2}{ (\delta\omega+ \delta)^2 + 4 \gamma_f^2}.
\end{align}
Averaging over all orientations, we obtain
\begin{align}
\int \textrm{Prob}^\textrm{same} (s) \; s \; ds &=
 \sqrt{\frac{2}{3}} 
\cdot
\frac{2}{3\sqrt{3}}
\cdot \sqrt{ \frac{ 2\gamma_f^2}{ \delta\omega^2 + 4 \gamma_f^2} }\\
\int \textrm{Prob}^\textrm{diff} (s) \; s \; ds &\simeq
 \sqrt{\frac{2}{3}} 
\times 0.6507
\times
 \sqrt{ \frac{ 2\gamma_f^2}{ (\delta\omega+\delta)^2 + 4 \gamma_f^2} },
\end{align}
where the middle factors arise from angular averaging of the matrix elements $g$ and $h$ of dipolar interactions.
Interestingly, the angle-averaged matrix element of flip-flop interaction is slightly larger for inter-group interaction than for intra-group interaction $0.6507 > 2/3\sqrt{3}$,
which explains the fact that the depolarization rate at the two-group resonance $\delta =0$ is slightly larger than four times that of a single group. 
We note that for the theory curves presented in the main text we further average over the energy mismatch $\delta \omega$ arising from inhomogeneous broadening, which we model using a Gaussian distribution with full width at half maximum (FWHM) $W\sim (2\pi)~9$~MHz.

\subsubsection*{Extension of the Spin Lifetime via Spin-locking}
Under strong driving conditions of a spin-locking sequence, the preferred quantization axes of spins and fluctuators are re-defined by the microwave driving. Specifically, in the rotating frame the eigenstates of a single particle Hamiltonian become $\ket{\pm} = (\ket{m_s=0}\pm\ket{m_s=-1})/\sqrt{2}$ and $\ket{m_s=+1}$ with corresponding energy eigenvalues $\pm\Omega/2$ and $0$ (up to on-site disorder due to the inhomogeneous broadening). 
Interestingly, the intra-group flip-flop interactions in this basis are strongly suppressed:
\begin{align}
\bra{+,-} H_{dd} \ket{-,+} & = 0\\
\bra{\pm,m_s=+1} H_{dd} \ket{m_s=+1,\pm} &=-J_0/r^3 ( g/2),
\end{align}
which implies that the spin lifetime $T_1^\rho$ along the $\ket{\pm}$ states is limited by resonant exchange with the third state $\ket{m_s=+1}$. The matrix element of such a process is suppressed by a factor of 2, which, together with the three level nature of the rate model in Eq.~\eqref{eq:pop1_exp}, leads to a factor of $2^2 \times 3=12$ improvement for $T_1^\rho$ compared to $T_1$.
For the theory curves presented in the main text (Fig.~4), we also include the effects of off-resonant interactions as well as interactions between different groups as in the previous section. 

\section{Charge Diffusion Model}
\begin{figure}[t]
\includegraphics[width=1\textwidth]{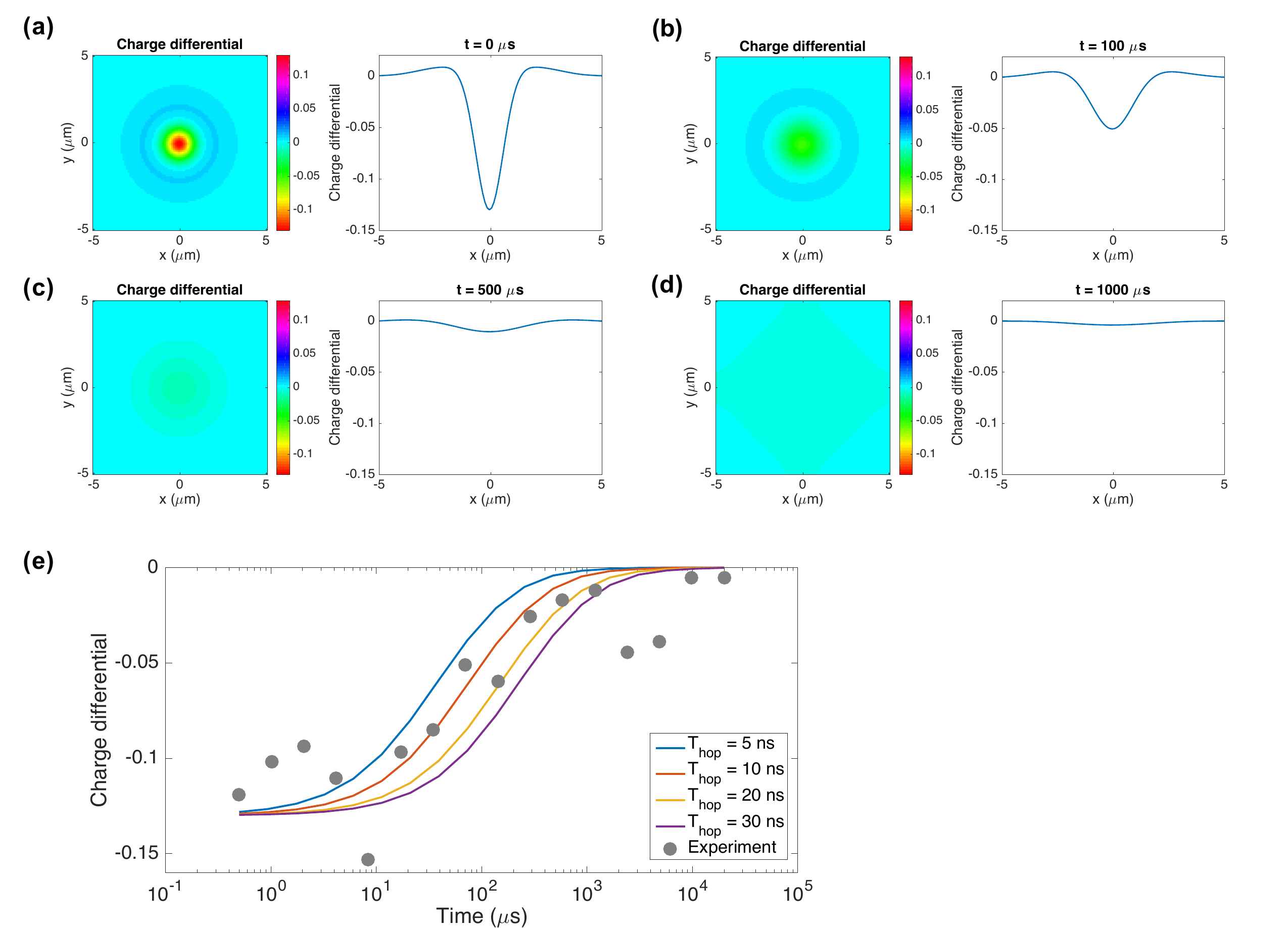} 
\caption{\textbf{Charge-state Dynamics.} 
\textbf{(a)-(d)} Simulated two-dimensional charge distribution (relative increase of NV$^-$ population compared to equilibrium) at different times after initialization, (a) $\tau$ = 0, (b) $\tau$ = 100, (c) $\tau$ = 500 and (d) $\tau$ = 1000 $\mu s$. We assumed an electron hopping timescale, $T_{hop}$ = 10~ns, and typical hopping distance, $a$ = 5~nm.
\textbf{(e)} Charge differential at the center measured over time (grey data). Colored solid lines indicate the diffusion simulation results calculated for different hopping times $T_{hop}$. 
}  \label{fig:charge}
\end{figure}

To model the observed charge state dynamics, we consider a classical diffusion equation, $\partial_t \delta n = D (\partial_{xx} + \partial_{yy}) \delta n$, where $D$ is the diffusion constant and $\delta n$ is the normalized excess or depletion of the NV$^-$ charge state of NV centers compared to the equilibrium value (charge differential).
We note that these charge diffusion experiments have been performed on a bulk piece of diamond under confocal excitation. However, due to the increased size of the confocal excitation spot in the direction normal to the diamond surface, we assume an effective 2D system.
For a local electron tunneling process, the diffusion constant can be expressed as $D = a^2/T_{hop}$, where $a\sim 5$~nm is the typical NV separation, and $T_{hop}$ is the electron hopping time.
Figure \ref{fig:charge}a-d summarizes the prediction of the diffusion model for $T_{hop}=10$~ns as a function of time, after the system is initialized into an out-of-equilibrium charge state (modeled after Fig.~4c main text). 
We calculate the expected rates of charge recovery at the center for various hopping rates, showing good agreement with the observed data when $T_{hop}\sim10$~ns.

 \newpage
\bibliography{depol_supp_bibtex}